\documentclass[aps,prl,twocolumn,superscriptaddress]{revtex4}
\usepackage{color}
\usepackage{graphicx}
\usepackage{amsmath}

\begin{document}
\title{Toroidal qubits: naturally-decoupled quiet artificial atoms}

\author{Alexandre M. Zagoskin}
\affiliation{Physics Department, Loughborough University, Loughborough LE11 3TU, United Kingdom}
\affiliation{iTHES Research Group, RIKEN, Saitama 351-0198, Japan}
\author{Arkadi Chipouline}
\affiliation{Institute of Applied Physics, Abbe Center of Photonics,
Friedrich-Schiller-Universit\"{a}t Jena, Max-Wien-Platz 1, 07743 Jena, Germany}
\author{Evgeni Il'ichev}
\affiliation{Leibniz Institute of Photonic Technology, P.O. Box 100239, D-07702 Jena, Germany}
\author{J. Robert Johansson}
\affiliation{iTHES Research Group, RIKEN, Saitama 351-0198, Japan}
\author{Franco Nori}
\affiliation{Center for Emergent Matter Science, RIKEN, Saitama 351-0198, Japan}
\affiliation{Physics Department, The University of Michigan, Ann Arbor, Michigan, 48109-1040, USA}

\begin{abstract}
The requirements of quantum computations impose high demands on the level of qubit protection from perturbations; in particular, from those produced by the environment. Here we propose a superconducting flux qubit design that is naturally protected from ambient noise. This decoupling is due to the qubit interacting with the electromagnetic field only through its toroidal moment, which provides an unusual qubit-field interaction.
\end{abstract}

\maketitle


{\em Introduction.--} A key requirement for quantum computing hardware is a low enough decoherence rate, which would allow either the implementation of quantum error correction schemes \cite{quantum error correction}, or the operation of an adiabatic optimization process \cite{Sarandy-Lidar}. Despite significant recent improvements in their performance \cite{superconducting qubits performance}, superconducting qubits are still more vulnerable to decoherence produced by local (i.e., by the qubit itself) and ambient (originating from the environment, including the control and readout circuitry)  noise, than some other platforms, such as spin- and ion trap-based ones \cite{other platforms}. Nevertheless, scalability and well-developed fabrication techniques make superconductor-based implementations a very attractive option, both for adiabatic \cite{adiabatic-qc} and circuit-based \cite{Fowler et al} quantum computing. Various designs of ``quiet" or ``silent" superconducting qubits have been proposed (e.g., \cite{Silent qubits}), but they involve exotic superconductors and do not protect against the intrinsic low frequency noise originating from the qubit itself \cite{low-f-noise}. Here we propose a qubit design that is naturally insensitive to low-frequency noise and is well protected from other ambient noise sources, and therefore could be a good candidate for a superconducting qubit. This qubit is also interesting from the point of view of investigating interesting and largely unexplored phenomena on the interaction of an electromagnetic field with toroidal multipoles in the quantum regime.

\begin{figure}[t]
\includegraphics[width=1.0\columnwidth]{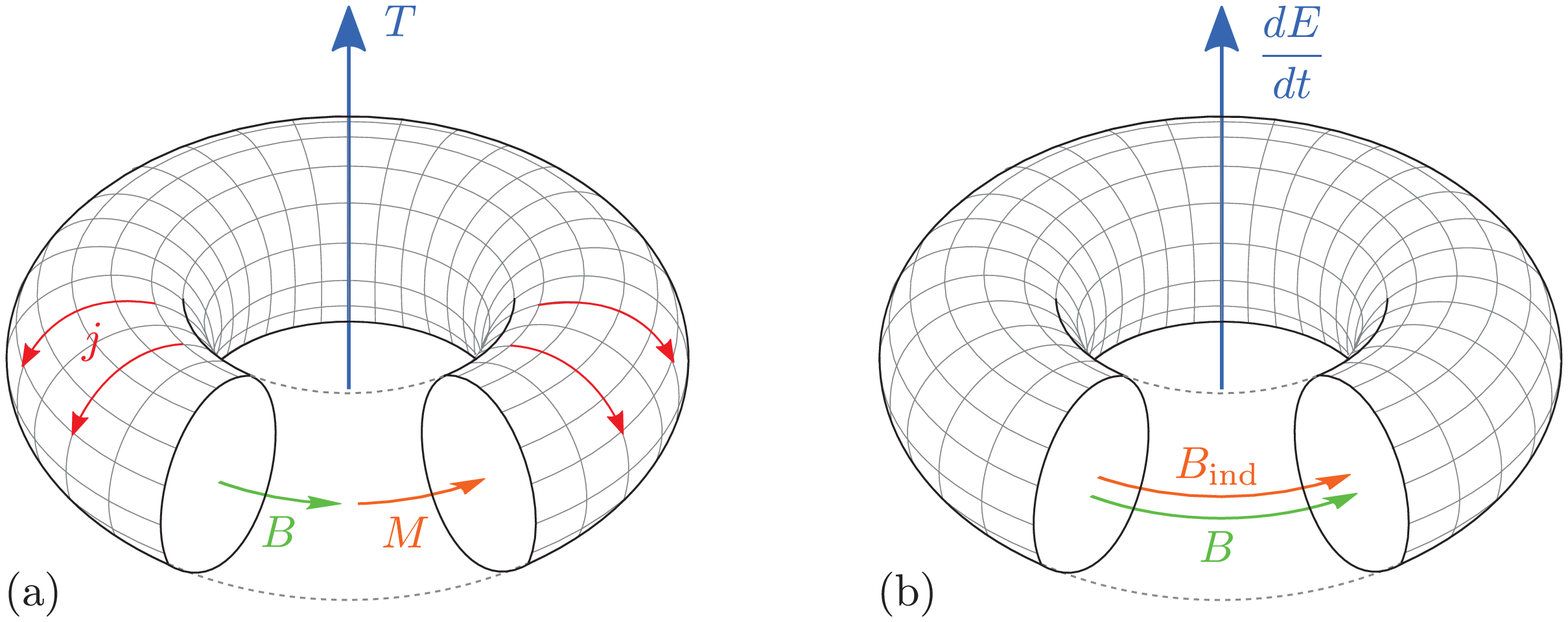}
\caption{(Color online) (a) Current $j$, magnetic field $B$, magnetic moment $M$ and dipolar toroidal moment $T$ of a toroidal coil. (b) Toroidal coil interacting with an external electric field.}
\label{fig:1}
\end{figure}

{\em Toroidal multipole moments.--}
The textbook multipole expansion of the electromagnetic field of a system of charges and currents routinely neglects a series of terms, which first appear in the higher orders of the expansion and that are independent of electric and magnetic multipoles. 
The toroidal multipoles were predicted by Zel'dovich in 1957 \cite{Zeld}; there he gave an example of the lowest-order (dipolar) toroidal moment, which 
 corresponds to the fields of a toroidal solenoid in the limit when its size tends to zero (Fig.~\ref{fig:1}a). The external magnetic field of such a structure is zero, but its interaction with an applied external magnetic field {\bf H} is nonzero and proportional to ${\bf T} \cdot \nabla\times{\bf H}$, where  $\bf T$ is the toroidal dipole moment (see below). It is part of the third-order expansion of the charge/current densities, so it appears in the usual expansions together with the octupole and magnetic quadrupole moments \cite{Dubovik90}. The toroidal moments are nontrivial objects, which have been considered mainly in nuclear physics, but did not attract too much attention in electrodynamics. Nevertheless, in the past years, numerous researchers are studying toroidal structures in optics and radio-frequencies \cite{Kaelberer,Teperik11,Fedotov13,Savinov13}; the toroidal ordering was observed in natural crystals (for a recent review see, e.g., Ref.~\cite{Kopaev}). 
 
Toroidal structures have interesting properties including: (1) {\em absence of generated fields}, for zero frequency, and the same for nonzero-frequency using anapoles, e.g. a combination of a toroid and a dipole; (2) {\em violation of the reciprocity theorem} \cite{Afanasiev01}; and (3) the anapole is also a potential candidate for dark matter in the universe \cite{Ho13}. Here we propose, for the first time to the best of our knowledge, the concept of quantum toroidal structures and show a direct application of these for quantum information processing.

Let us briefly recapitulate the properties of a dipolar toroidal moment (see, e.g., \cite{Afanasiev}). Consider the toroidal current distribution of Fig.~\ref{fig:1}a. The sheet current density {\bf J} produces the magnetization {\bf M} inside the torus:
\begin{equation}
{\bf J} = \nabla\times{\mathbf M}.
\label{eq:1}
\end{equation}
Since $\nabla\cdot{\bf M} = 0$, then
\begin{equation}
{\bf M} = \nabla\times{\mathbf T},
\label{eq:2}
\end{equation}
where {\bf T} is the dipolar toroidal moment. At large distances from the torus (i.e., in the limit when the diameter of the tube, and then the radius of the torus are taken to zero) the toroidal moment of the system remains finite and characterizes its electromagnetic potentials.

Remarkably, a toroidal moment couples to the {\em time} {\em derivatives} of the external electromagnetic field. In particular, for a  toroidal dipole in the limit of slow spatial variation of the external field, the coupling potential is \cite{Afanasiev}
\begin{equation}
U_{\rm int} = -(\mu_0\epsilon_0) (4\pi\epsilon_0)^{1/2}\left(\frac{d}{dt}{\bf E}_{\rm ext}\right)\cdot {\bf\bar{t}},
\label{eq:3}
\end{equation}
where
$\bar{\bf t}  = \int \!d^3r\:\, {\bf T}({\bf r}).$
For Fig.~\ref{fig:1}, with the torus axis directed along {\bf n}, diameter $D$, crossection $\pi R^2$, and considering the toroidal dipole as a toroidal solenoid with $N$ turns and current $I$ in each turn, we find \cite{Afanasiev}
\begin{equation}
\bar{\bf t}  = -2\pi^2 {\cal J}_0 {\bf n},\;\;\; {\cal J}_0  = \frac{NIR^2D}{4\pi (4\pi\epsilon_0)^{1/2}}.
\label{eq:5}
\end{equation}
Therefore, the coupling between the toroidal dipole and the electromagnetic field is given by
\begin{equation}
U_{\rm int} =  \frac{\mu_0\varepsilon_0 \pi N I R^2 D}{2}  \left(\frac{d{\bf E}_{\rm ext}}{dt}\right)\cdot {\bf n}.
\label{eq:6}
\end{equation}
The sign of this coupling is positive (i.e., if $d{\bf E}_{\rm ext}/dt$ is co-directional with $\bf \bar{t}$, then the energy of the system increases). This can be seen directly from the fourth Maxwell's equation:  The additional magnetic field ${\bf B}_{\rm ind}$ inside the torus, induced by the growing electric flux, will add to the field  ${\bf B} \propto {\bf M}$ (see Fig.~\ref{fig:1}b). Since the coupling to the external electric field is proportional to its time derivative, a dipole toroidal moment is {\it insensitive to low-frequency electric noise}, which is a major source of decoherence in quantum devices.

\begin{figure}%
\includegraphics[width=0.9\columnwidth]{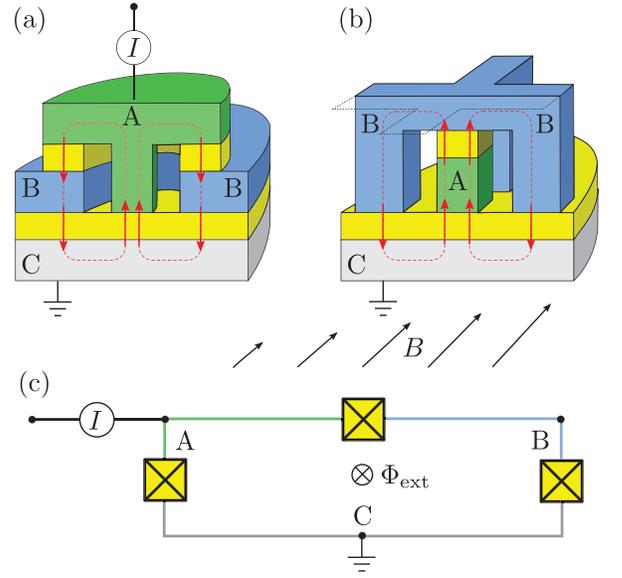}
\caption{(Color online) Toroidal qubit: (a) ``Closed" version; (b) ``Open" version. One of the two possible directions of the circulating Josephson currents is shown. (c) Equivalent circuit.}
\label{fig:2}%
\end{figure}

{\em Toroidal qubit.--}
We propose two possible designs of toroidal qubits (see Fig.~\ref{fig:2}). It can be seen from their lumped-elements circuit that these designs are topologically identical to one of the first successful superconducting qubit designs: The persistent current flux qubit \cite{Mooij}. Here, instead of a flat loop, the equilibrium currents flow along a completely, or partially, closed toroidal surface formed by the superconducting layers and tunneling barriers.

In the ``closed" version,  the superconducting layers completely enclose the internal volume, where the magnetic field generated by the Josephson currents is confined. The advantage of this design (the toroidal current flow and thus zero leakage of the magnetic field to the outside) is counterbalanced by the impossibility to bias the qubit to the vicinity of the degeneracy point by the external magnetic flux. Therefore, it is necessary to use a $\pi$-junction, and to fine-tune the qubit by the external bias current, as in the case of a phase qubit \cite{Martinis}. While qubits with $\pi$-junctions have been successfully demonstrated, their fabrication and incorporation in more complex designs remain a challenge \cite{Ryazanov}.

The ``open" version, which is similar to a classical implementation of a toroidal moment \cite{Kaelberer},  only approximates the toroidal current, but due to the holes in the electrode B, it can be tuned by the external magnetic field and does not require a $\pi$-junction. It has a gradiometric design, making it {\it less sensitive} to the ambient noise \cite{gradiometric} and to some extent compensates for the deviation from the ``ideal" toroidal design.

The Lagrangian of the system, as a function of $\Phi_A, \Phi_B, \dot{\Phi}_A$, and $\dot{\Phi}_B$, is given by (see Ref.~\cite{Zagoskin}, Ch.2):
\begin{eqnarray}
{\cal L}_{a,b} = \frac{C_F(\dot{\Phi}_A-\dot{\Phi}_B)^2}{2} +  \frac{C_A\dot{\Phi}_A^2}{2} +  \frac{C_B\dot{\Phi}_B^2}{2} +\nonumber\\ E_A\cos 2\pi\frac{\Phi_A}{\Phi_0} + E_B\cos 2\pi\frac{\Phi_B}{\Phi_0} +
 \Delta{\cal L}_{a,b},
\label{eq:7}
\end{eqnarray}
where
\begin{align}
 \Delta{\cal L}_{a} =& E_F \cos 2\pi\left(\frac{1}{2} + \frac{\Phi_A-\Phi_B}{\Phi_0}\right) + I_{\rm ext}(\Phi_A-\Phi_B), \nonumber\\
 \Delta{\cal L}_{b} =& E_F \cos 2\pi\left(\frac{\Phi_{\rm ext}}{\Phi_0} + \frac{\Phi_A-\Phi_B}{\Phi_0}\right)
\label{eq:8}
\end{align}
describe the potential (Josephson) energy. The sub-indices $a$ and $b$ refer to the toroidal qubits shown in $(a)$ and $(b)$, respectively, in Fig.~\ref{fig:2}.
The variables
\begin{equation}
\Phi_{\alpha} = \int^{t} V_{\alpha}(t') dt',\;\;\;\; \alpha = \{A,B\},
\label{eq:8prime}
\end{equation}
are related to the voltages at the corresponding nodes, measured with respect to the ground node (which can be chosen arbitrarily), and $\Phi_0 = h/2e$ is the magnetic flux quantum. Finally, $I_{\rm ext}$ and $\Phi_{\rm ext}$ are the external tuning parameters (bias current and the magnetic flux through the corresponding loop, respectively).

To decouple the time derivatives in the Lagrangian, we introduce new variables
\begin{eqnarray}
\Phi_A = \psi_A \cos\Theta + \psi_B \sin\Theta, \nonumber\\
\Phi_B = -\psi_A \sin\Theta + \psi_B \cos\Theta, \\
\cos\Theta = \frac{1}{2}\sqrt{2+\frac{C_A-C_B}{\Delta C}}, \nonumber \\
\sin\Theta = \frac{1}{2}\sqrt{2-\frac{C_A-C_B}{\Delta C}}, \nonumber\\
\Delta C = \frac{1}{2}\sqrt{(C_A-C_B)^2+4C_F^2}.
\label{eq:9}
\end{eqnarray}
Then the  Lagrangian becomes
\[
{\cal L} = \frac{1}{2} \left( (\bar{C}+\Delta C)\dot{\psi}_A^2 + (\bar{C}-\Delta C)\dot{\psi}_B^2 \right) + \dots,
\]
where $\bar{C} = C_F + (C_A+C_B)/2$. We can introduce the canonical momenta (``generalized electric charges")
\begin{align}
p_A &= (\bar{C}+\Delta C)\dot{\psi}_A, \\
p_B &= (\bar{C}-\Delta C)\dot{\psi}_B,
\label{eq:10}
\end{align}
and the Hamiltonian then becomes
\begin{eqnarray}
H_{a,b} &=& \frac{p_A^2}{2(\bar{C}+\Delta C)} + \frac{p_A^2}{2(\bar{C}-\Delta C)} \nonumber\\
&+& U(\psi_A,\psi_B)+\Delta U_{a,b}(\psi_A,\psi_B).
\label{eq:11}
\end{eqnarray}
The potential energy terms in Eq.~(\ref{eq:11}) are:
\begin{widetext}
\begin{eqnarray}
&& U(\psi_A,\psi_B) = -E_A\cos\left(2\pi\frac{\psi_A \cos\Theta + \psi_B \sin\Theta}{\Phi_0}\right)-E_B\cos\left(2\pi\frac{-\psi_A \sin\Theta + \psi_B \cos\Theta}{\Phi_0}\right),\;\;\; \\
\Delta U_a \!\!\!&=&\!\!\! -E_F\cos 2\pi\left(\frac{1}{2}+\frac{\psi_A (\cos\Theta+\sin\Theta) + \psi_B (\sin\Theta-\cos\Theta)}{\Phi_0}\right) + 
I_{\rm ext}\left[\psi_A (\cos\Theta+\sin\Theta) + \psi_B (\sin\Theta-\cos\Theta)\right],\;\;\;\;\;\;\\
\Delta U_{b} \!\!\!&=&\!\!\! -E_F\cos 2\pi\left(\frac{\Phi_{\rm ext}}{\Phi_0}+\frac{\psi_A (\cos\Theta+\sin\Theta) + \psi_B (\sin\Theta-\cos\Theta)}{\Phi_0}\right).\;\;\;
\label{eq:12}
\end{eqnarray}
\end{widetext}
The electric charge on a given node equals
\begin{eqnarray*}
Q_{A,B} = \frac{\partial{\cal L}}{\partial\dot{\Phi}_{A,B}} = p_A\frac{\partial\dot{\psi}_A}{\partial\dot{\Phi}_{A,B}} + p_B\frac{\partial\dot{\psi}_B}{\partial\dot{\Phi}_{A,B}},
\end{eqnarray*}
that is,
\begin{eqnarray}
Q_A = p_A \cos\Theta + p_B \sin\Theta, \nonumber\\
Q_B = -p_A \sin\Theta + p_B \cos\Theta.
\label{eq:13}
\end{eqnarray}
This allows us to estimate the electric dipole moment of the qubit. In the case of a realistic choice of parameters (see below) this moment is negligibly small. This is to be expected, since the original persistent current qubit was specifically designed to minimize the influence of the electric charge noise \cite{Mooij}. The toroidal moment of the qubit is determined by the Josephson current flow pattern through the coefficient ${\cal J}_0$ in Eq.~(\ref{eq:5}), which now becomes proportional to the current operator. We can approximately express it as
\begin{equation}
\hat{\bf t} = - 2\pi^2\hat{\cal J}_0{\bf n} \approx \frac{1}{4\pi^3}V_{\rm eff} \hat{I},
\label{eq:14}
\end{equation}
where $\hat{I}$ is the operator of the Josephson current flowing between the electrodes A and B (that is, circulating around the qubit loop - see the equivalent scheme), and $V_{\rm eff}$ is the effective volume encased by the current (in the case of a torus, Fig.~\ref{fig:1}, $V_{\rm eff} = \pi^2DR^2$).

{\em Qubit-field coupling strength.--}
The corresponding coupling between the qubit and the external electric field is then
\begin{eqnarray}
U_{\rm int} \approx \frac{\mu_0\varepsilon_0 V_{\rm eff}}{2\pi} \langle \hat{I} \rangle  \left(\frac{d{\bf E}_{\rm ext}}{dt}\right)\cdot {\bf n}.
\label{eq:15}
\end{eqnarray}
Estimating $V_{\rm eff} \sim \left(10\;\mu{\rm m}\right)^3$ and $\langle \hat{I} \rangle \sim 1\;\mu{\rm A}$, we see that
\begin{equation}
U_{\rm int}({\rm J}) \sim 2 \times 10^{-38} \;\frac{dE}{dt} \;\;({\rm V}\cdot({\rm m{\cdot}s})^{-1}).
\label{eq:16}
\end{equation}
For a field with amplitude 100 kV/m and frequency 100 GHz this yields the interaction strength $\sim 1.5 \times 10^{-23}$~J, or $\sim 20$ GHz.

\begin{figure}[t]
\includegraphics[width=0.9\columnwidth]{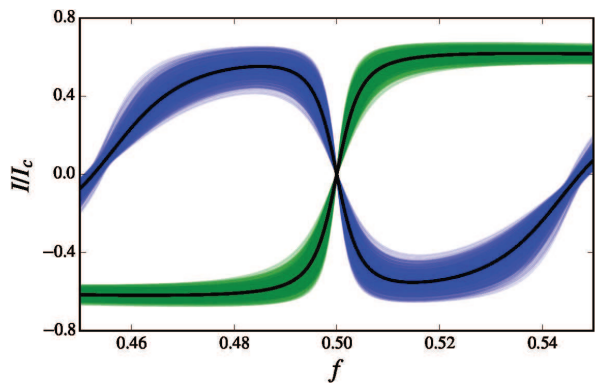}
\caption{(Color online) The circulating current $I$ for the ground state (green) and the first excited state (blue) of the flux-biased toroidal qubit, versus the reduced magnetic flux $f = \Phi_{\rm ext}/\Phi_0$. Here we use the parameters $C_A/\bar{C} = C_B/\bar{C} = 1$, $C_F/\bar{C} = 0.5$, $E_A = E_B = E_J$, $E_F = 0.8 E_J$, and $E_{\bar{C}}/E_J = 1/40$, where $E_{\bar{C}} = (2e)^2/(2\bar{C})$ is the average charging energy and $E_J$ the Josephson energy corresponding to the critical current $I_c$. Typical values of the critical current and charging energy considered here are $I_c \sim 1\;\mu$A and $\bar{C} \sim 1$ fF. The thin curves represent the critical current $I$ for 1000 random realization that include a 10\% disorder in $C_A, C_B, C_F, E_A, E_B$ and $E_F$. We note that the qubit is stable with respect to moderate variations in these parameters.}
\label{fig:I-vs-f-noisy}
\end{figure}

The toroidal moment of the qubit depends on its quantum state. In the ``physical" basis of states $|L\rangle$ and $|R\rangle$ (i.e., those with the Josephson currents flowing like in Fig.~\ref{fig:2}b or in the opposite direction) it can be written as
\begin{equation}
\hat{\bf t} = - 2\pi^2|J_0|{\bf n} \sigma_z.
\label{eq:16a}
\end{equation}
The Josephson currents for the two lowest-energy states of the external-flux-biased qubit design, as a function of the reduced magnetic flux $f$, is shown in Fig.~\ref{fig:I-vs-f-noisy}.
The effective qubit Hamiltonian can be obtained in a standard way by quantizing Eq.~(\ref{eq:11}) and considering only the subspace spanned by the two lowest-lying states \cite{Mooij}:
\begin{equation}
H_{\rm qb} = -\frac{\hbar}{2}(\Delta \sigma_x + \epsilon \sigma_z).
\label{eq:16b}
\end{equation}
The bias $\epsilon$ is controlled by the parameters $I_{\rm ext}$ and $\Phi_{\rm ext}$, while the tunneling splitting $\Delta$ is determined by the ratio of charging and Josephson energies of the junctions (see Refs.~\cite{Mooij,Zagoskin}) and is typically in the GHz range. 

If the electric field, with which the qubit interacts, is parallel to the $z$ axis, then the field-qubit interaction term in the Hamiltonian is
\begin{equation}
H_{\rm int} \approx \hbar\lambda \left(\frac{dE_{\rm ext}}{dt}\right) \sigma_z,
\label{eq:16c}
\end{equation}
where
\begin{equation}
\lambda = \frac{\mu_0\varepsilon_0 V_{\rm eff} }{2\pi} I_J  \left(\frac{dE_{\rm ext}}{dt}\right).
\label{eq:17}
\end{equation}

{\em Conclusions.--}
The closed toroidal qubit is protected from the ambient low-frequency noise (e.g., 1/f-noise). Its reaction to high-frequency ambient noise is less important, since it is routinely filtered out in standard experimental setups. The open toroidal qubit is also well protected, partly due to its gradiometric design. As we have shown, the numerical calculations with reasonable parameters also show tolerance of the system to the parameters dispersion. 

Despite looking exotic, toroidal qubits can be fabricated using current modern superconducting technology \cite{Andres,Tolpygo}. For example, the structure (a) of Fig.~2 is formed from the stack containing two Josephson junction in series. By making use of electron beam lithography and etching, the inner dimension of the torus and the outer dimensions of the junction can be shaped simultaneously.  Depositing dielectric and lift-off consequently will result in a qubit of type (a) without the upper electrode. After a planarization of the structure this electrode can be deposited on top, thus completing the fabrication process. Fabrication of the control and readout circuitry will not present any challenges beyond the routine. 

Due to its effective decoupling from the environment, only the decoherence sources inside the toroidal section of the qubit limit the qubit decoherence time.
The material properties of the dielectric inside the torus thus acquire the key importance. This presents both a challenge and an opportunity. We know that a drastic increase of decoherence can be achieved by improving the quality of the tunneling barrier in a qubit \cite{Martinis05}. However, the insensitivity of the toroidal design to the external noise would make it a good tool for the investigation of low-temperature noise properties of different dielectrics for microwave quantum engineering. 

AC acknowledges financial support by the Federal Ministry of Education and Research in the frame of a PhoNa Project.
EI acknowledges funding from the European Community's Seventh Framework Programme (FP7/2007-2013) under Grant No. 270843 (iQIT).
FN is partially supported by the RIKEN iTHES Project, MURI Center for Dynamic Magneto-Optics, and a Grant-in-Aid for Scientific Research (S).

\end{document}